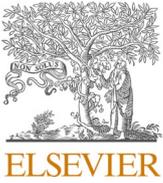
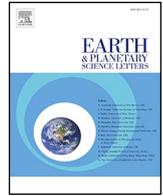

# Condensate evolution in the solar nebula inferred from combined Cr, Ti, and O isotope analyses of amoeboid olivine aggregates

Christian A. Jansen [a], Christoph Burkhardt [a,b,*], Yves Marrocchi [c], Jonas M. Schneider [a,b], Elias Wölfer [a,b], Thorsten Kleine [a,b]

[a] *Institut für Planetologie, University of Münster, Wilhelm-Klemm-Straße 10, Münster D-48149, Germany*
[b] *Max Planck Institute for Solar System Research, Justus-von-Liebig-Weg 3, Göttingen D-37077, Germany*
[c] *Centre de recherches pétrographiques et géochimiques (CRPG), CNRS, UMR 7358, F-54000, Nancy, France*

## A B S T R A C T

Refractory inclusions in chondritic meteorites, namely amoeboid olivine aggregates (AOAs) and Ca-Al-rich inclusions (CAIs), are among the first solids to have formed in the solar system. The isotopic composition of CAIs is distinct from bulk meteorites, which either results from extreme processing of presolar carriers in the CAI-forming region, or reflects an inherited heterogeneity from the Sun's parental molecular cloud. Amoeboid olivine aggregates are less refractory than CAIs and provide a record of how the isotopic composition of solid material in the disk may have changed in time and space. However, the isotopic composition of AOAs and how this composition relates to that of CAIs and later-formed solids is unknown. Here, using new O, Ti, and Cr isotopic data for eight AOAs from the Allende CV3 chondrite, we show that CAIs and AOAs share a common isotopic composition, indicating a close genetic link and formation from the same isotopic reservoir. Because AOAs are less refractory than CAIs, this observation is difficult to reconcile with a thermal processing origin of the isotope anomalies. Instead, the common isotopic composition of CAIs and AOAs is readily accounted for in a model in which the isotopic composition of infalling material from the Sun's parental molecular cloud changed over time. In this model, CAIs and AOAs record the isotopic composition of the early infall, while later-formed solids contain a larger fraction of the later, isotopically distinct infall. This model implies that CAIs and AOAs record the isotopic composition of the Sun and suggests that the nucleosynthetic isotope heterogeneity of the solar system is predominantly produced by mixing of solar nebula condensates, which acquired their distinct isotopic compositions as a result of time-varied infall from the protosolar cloud.

## 1. Introduction

Chondritic meteorites derive from parent bodies that did not undergo melting and chemical differentiation, and thus provide a record of the materials present in the solar nebula at different times and places of protoplanetary disk evolution. As such, the investigation of the distinct components preserved in chondrites—refractory inclusions, chondrules, metal, and matrix—provides some of the most direct constraints on the physicochemical conditions under which solid material in the disk was formed, transported, and mixed. To this end, refractory inclusions are of considerable interest, because they are the oldest dated samples of the solar system and thus provide a record of evaporation and condensation processes at high ambient temperatures close to the young Sun (e.g., Krot, 2019). In addition, as is evident from their ubiquitous presence in carbonaceous chondrites and even comets (McKeegan et al., 2006; Matzel et al., 2010), refractory inclusions bear testimony to large-scale transport of solid material through the disk, from their formation region close to the Sun to the outer solar system, where carbonaceous chondrites and comets later formed (e.g., Ciesla, 2007).

Refractory inclusions are divided into two types: Ca-Al-rich inclusions (CAIs) and amoeboid olivine aggregates (AOAs). The chemical and mineralogical composition of CAIs is consistent with them being the first solids that formed in a cooling gas of solar composition (Grossman, 1972). Consistent with this, CAIs are the oldest dated objects of the solar system with an absolute age of between ~4567 and ~4568 Ma (Connelly et al., 2012; Amelin et al., 2010; Bouvier and Wadhwa, 2010), and are commonly used to define the onset of solar system formation in cosmochemistry (Brennecka et al., 2020). The isotopic composition of CAIs is distinct from that of subsequently formed solar nebula solids, where apart from different O isotope compositions (Clayton et al., 1973), CAIs are characterized by some of the largest nucleosynthetic isotope anomalies observed among materials that formed in the solar system (see summary in Dauphas and Schauble, 2016). These anomalies arise from the heterogeneous distribution of presolar material and, as such, reflect an inherent isotopic difference in the precursor material of a sample. Nucleosynthetic isotope anomalies, therefore, provide information on genetic links among solid materials of the disk. For instance, these anomalies show that bulk meteorites can be assigned to either the non-carbonaceous (NC) and carbonaceous (CC) group (Warren, 2011; Budde et al., 2016), which have been suggested to represent material from the inner and outer solar system, respectively (Kruijer et al., 2017; Warren, 2011). Of note, CC meteorites generally have isotope compositions intermediate between those of NC meteorites and CAIs, implying that compared to the NC reservoir, the CC reservoir contains a greater proportion of material with a CAI-like isotopic composition (Nanne et al., 2019; Burkhardt et al., 2019; Render et al., 2022). However, the origin of the isotope anomalies in CAIs themselves, and of the isotopic difference between CAIs and bulk meteorites is debated. These anomalies may either reflect extreme thermal processing in the disk during which specific isotopically anomalous presolar phases were preferentially incorporated into the CAI-forming gas (Dauphas et al., 2004; Paton et al., 2013; Trinquier et al., 2009), or they are inherited from the Sun's parental cloud and reflect a temporal change in the isotopic composition of matter added to the disk during collapse of the protosolar cloud (Burkhardt et al., 2019; Jacquet et al., 2019; Nanne et al., 2019).

Amoeboid olivine aggregates are irregularly-shaped, sub-mm to mm-sized objects which predominantly consist of forsteritic (Fo$_{>98}$) olivine with inclusions of fine-grained aggregates of CAI-like material and minor Fe-Ni metal (Komatsu et al., 2001; Ruzicka et al., 2012; Jacquet and Marrocchi, 2017). They are commonly interpreted as aggregates of relatively fine-grained condensates and provide evidence for gas-solid reactions at temperatures below the stability field of CAIs (Grossman and Steele, 1976; Sugiura et al., 2009; Krot et al., 2004). As such, AOAs can be considered to represent an intermediate stage of condensate evolution in the solar nebula, between early high-temperature condensates (CAIs) and less-refractory, later-formed solids like chondrules or their precursors (McSween, 1977; Krot et al., 2004; Ruzicka et al., 2012). This makes AOAs uniquely useful to study how the isotopic composition of solar nebula condensates evolved, and how this relates to the isotopic difference between CAIs and later-formed solids such as for example chondrules.

Several studies have shown that AOAs and CAIs share the same $^{16}$O-rich isotope composition, suggesting that both objects are genetically linked (Hiyagon and Hashimoto, 1999; Krot et al., 2004). The O isotope signatures primarily reflect gas-solid or gas-melt interactions during formation of these objects, and so indicate that CAIs and AOAs formed from a gas of similar O isotope composition. As such, O isotopes do not provide a direct record of potential genetic links between the solid precursor material of CAIs and AOAs. Instead, nucleosynthetic isotope anomalies are more suitable to identify such genetic links, but to date there is only a single AOA which has been investigated for its nucleosynthetic Ti and Cr isotope signature (Trinquier et al., 2009). Although this sample is isotopically indistinguishable from CAIs, based on a single sample it is not possible to assess the genetic link between CAIs and AOAs in a comprehensive manner.

Here we report combined O, Ti, and Cr isotopic data for eight AOAs from the CV3 carbonaceous chondrite Allende. These data are used to assess genetic links between CAIs and AOAs, which in turn has important implications for understanding the origin of the nucleosynthetic isotope heterogeneity of the solar system and allows reconstructing the isotopic evolution of solar nebula condensates over time.

## 2. Samples and methods

Eight AOAs from the Allende CV3 chondrite were selected for this study. They were identified and characterized by scanning electron microscopy (SEM) in polished slices of the Allende chondrite (Fig. S1; Supplementary Information). The selected AOAs have apparent diameters of between 3.0 and 6.4 mm and consist mainly of forsteritic olivine (Fo$_{>90}$), Ca-Al-rich assemblages (Al-rich diopside + anorthite + spinel), and minor Fe-Ni-metal. Common secondary minerals include ferrous olivine (Fo$_{<60-90}$), Ca-Fe-rich pyroxenes, Fe-Ni-sulfides, nepheline, and sodalite. Many forsterite grain boundaries show 120° triple junctions, indicative of high-temperature nebular annealing after condensation (Han and Brearley, 2015; Komatsu et al., 2001). Along the (sub)grain boundaries, forsterite is often replaced to some extent by ferrous olivine, reflecting secondary grain growth or Fe-Mg exchange. All AOAs of this study consist of fine-grained aggregates surrounded by granular olivine, and can thus be classified as the "rimmed type" (Kornacki and Wood, 1985a, 1985b). Among the AOAs of this study, AOA-17 is special, because it contains two relict CAI cores of ~1 mm in apparent diameter.

Two samples (AOA-1 and AOA-5) were removed manually from the slices using a diamond saw, followed by hand picking under a stereoscopic microscope. The six other AOAs were extracted using a New Wave Research MicroMill equipped with diamond-coated 2 mm hollow drills (Figs. S2, S3; Supplementary Information). The drilling depth was approximately 0.5–1 mm depending on each AOA's apparent size. In most cases, the AOA material was too brittle to extract a solid drill core, and instead the powder created during the drilling process was collected. Only for AOA-22 a solid drill core was obtained, which was processed separately from the powder of this sample. To assess whether the AOA samples were contaminated with adjacent material during drilling, the sample powders were inspected by eye and with a binocular microscope, and back-scatter electron (BSE) images were taken before and optical images after drilling (Fig. S3; Supplementary Information). On this basis, the proportion of AOA material in the analyzed samples is estimated to be >90 % for most samples, although AOA-10 and AOA-23 may contain larger fractions of surrounding matrix material of up to ~50 %. All samples (0.6–12.0 mg) were weighed to high precision by microbalance before digestion. A representative piece of each AOA was preserved and mounted in epoxy resin for O isotope analyses by secondary ion mass spectrometry (SIMS).

The O isotopic measurements were performed using a CAMECA IMS 1270 E7 at CRPG-CNRS (Bouden et al., 2021). $^{16}$O$^-$, $^{17}$O$^-$, and $^{18}$O$^-$ ions produced by a Cs$^+$ primary ion beam (~4 μm, 500 pA) were measured in multi-collection mode using two off-axis Faraday cups for $^{16,18}$O$^-$ and the axial electron multiplier (EM) for $^{17}$O$^-$. To remove $^{16}$OH$^-$ interference on the $^{17}$O$^-$ peak and achieve maximum flatness atop the $^{16}$O$^-$ and $^{18}$O$^-$ peaks, the entrance and exit slits of the central EM were adjusted to achieve a mass resolving power m/Δm of ~6000 for $^{17}$O$^-$. The Faraday cups were set on exit slit 1 (m/Δm = 2500). The total measurement duration was 260 s, comprising 60 s of pre-sputtering and 200 s of measurement. The terrestrial standards San Carlos olivine, Ipanko spinel, and Gold enstatite were used to define the instrumental mass fractionation (IMF) line for the three O isotopes and correct for IMF due to matrix effects in olivine. Oxygen isotopic compositions are reported in the δ-notation relative to terrestrial standard mean ocean water [δ$^i$O = ($^i$O/$^{16}$O)$_{sample}$/($^i$O/$^{16}$O)$_{SMOW}$− 1) × 10$^3$ where $i$ = 17, 18], and as displacement from the terrestrial fractionation line, i.e., Δ$^{17}$O = δ$^{17}$O − 0.52 × δ$^{18}$O. The uncertainties on Δ$^{17}$O were calculated by quadratically summing the errors on δ$^{17}$O and δ$^{18}$O and the standard deviation (s.d.) of the Δ$^{17}$O values obtained for terrestrial standards.

For chemical, and Ti and Cr isotopic analyses the AOAs were digested using concentrated HF/HNO$_3$ (3:1) in Parr bombs at 190 °C for 96 h. After complete digestion, a small aliquot (0.5–1 %) was removed from the sample solution and used for concentration measurements of selected major and trace elements (Table 1) using a ThermoScientific

**Table 1**
Textural and compositional data for Allende AOAs and USGS reference material BHVO-2.

| Sample | AOA-1 | AOA-5 | AOA-10 | AOA-17 | AOA-18 | AOA-20 | AOA-22c | AOA-22p | AOA-23 | BHVO-2 |
|---|---|---|---|---|---|---|---|---|---|---|
| Apparent dimension [mm] | 6.2 × 4.2 | 5.2 × 3.1 | 4.1 × 3.4 | 6.4 × 3.6 | 5.8 × 4.2 | 5.7 × 3.7 | 6.1 × 3.9 | | 3.0 × 2.9 | |
| Texture | rimmed | rimmed | rimmed | rimmed | rimmed | rimmed | rimmed | rimmed | rimmed | |
| Mineralogy | ol, aug, spl, met, sf, nph, px | ol, di, chr, rt, met, sf, nph, px | ol, met, sf, px | ol, di, met, sf, px, relict CAIs | ol, aug/di, met, sf, px ± an, spl, nph, sdl | | | | | |
| Separation | handpicked | handpicked | drilled | drilled | drilled | drilled | drilled | drilled | drilled | |
| Digested mass [mg] | 11.951 | 4.130 | 1.571 | 3.373 | 11.854 | 2.738 | 0.644 | 1.694 | 3.642 | |
| Element composition [µg/g] | | | | | | | | | | |
| Na | 9086 | 11,351 | 2702 | 19,871 | 2183 | 7826 | 12,570 | 11,017 | 4135 | 17,088 |
| Mg | 231,227 | 214,645 | 223,019 | 161,622 | 243,011 | 214,227 | 210,117 | 225,404 | 221,019 | 50,955 |
| Al | 20,811 | 22,428 | 11,962 | 57,041 | 13,027 | 21,012 | 21,243 | 17,374 | 13,316 | 73,753 |
| Ca | 19,295 | 21,747 | 12,060 | 49,943 | 12,365 | 21,098 | 20,190 | 16,382 | 13,375 | 66,057 |
| Sc | 16 | 21 | | 47 | 9 | 17 | | 21 | 12 | 32 |
| Ti | 1095 | 1201 | 600 | 3189 | 629 | 1012 | 1152 | 988 | 712 | 16,963 |
| V | 106 | 91 | | 340 | 66 | 95 | | 76 | 57 | 358 |
| Cr | 1721 | 1470 | 2037 | 1182 | 1904 | 1760 | 897 | 1046 | 1768 | 272 |
| Fe | 171,863 | 191,930 | 215,202 | 120,964 | 169,842 | 170,358 | 149,593 | 160,596 | 177,754 | 102,356 |
| Mn | 1264 | 1341 | 1366 | 949 | 917 | 1374 | 796 | 907 | 958 | 1376 |
| Co | 118 | 130 | 314 | 108 | 301 | 208 | 69 | 199 | 329 | 41 |
| Ni | 2354 | 2188 | 8579 | 3186 | 7932 | 6250 | 1114 | 6798 | 9392 | 130 |
| Cu | 9 | 10 | 37 | 32 | 44 | 32 | 20 | 31 | 52 | 220 |
| Zn | 78 | 73 | 168 | 196 | 114 | 106 | 292 | 126 | 99 | 128 |
| Rb | 3.6 | 4.4 | 1.8 | 6.9 | 1.0 | 2.6 | 7.2 | 4.3 | 1.8 | 8.7 |
| Sr | 17 | 19 | 15 | 49 | 13 | 19 | 24 | 19 | 13 | 409 |
| Y | 3.1 | 3.9 | 1.6 | 9.5 | 1.7 | 3.3 | 3.4 | 2.8 | 1.5 | 22 |
| Zr | 9.6 | 12.1 | 5.2 | 27.8 | 5.2 | 11.2 | 10.3 | 8.9 | 5.3 | 173 |
| Nb | 0.78 | 0.83 | 0.92 | 2.19 | 0.59 | 1.05 | 1.76 | 0.77 | 0.62 | 18 |
| Mo | 1.10 | 1.31 | | | | | | | | 2.3 |
| Ru | 0.48 | 1.06 | | 2.85 | 0.59 | 0.69 | | | 0.59 | |
| Cd | 0.31 | 0.38 | 0.08 | 1.34 | 0.12 | 0.26 | 0.73 | 0.59 | 0.21 | 0.20 |
| Ba | 8.4 | 8.4 | 16.9 | 19.9 | 4.9 | 9.7 | 67.8 | 5.8 | 4.9 | 135 |
| La | 0.54 | 0.54 | 0.28 | 1.6 | 0.33 | 0.65 | 0.74 | 0.64 | 0.36 | 14 |
| Ce | 1.3 | 1.3 | 0.69 | 3.8 | 0.80 | 1.2 | 1.3 | 1.4 | 0.81 | 34 |
| Pr | 0.20 | 0.22 | 0.12 | 0.58 | 0.11 | 0.26 | 0.25 | 0.22 | 0.12 | 4.9 |
| Nd | 0.91 | 1.00 | 0.61 | 2.70 | 0.54 | 1.13 | 1.08 | 0.86 | 0.52 | 20 |
| Sm | 0.34 | 0.35 | 0.15 | 1.03 | 0.15 | 0.47 | 0.40 | 0.28 | 0.20 | 5.5 |
| Eu | 0.15 | 0.19 | 0.09 | 0.42 | 0.11 | 0.16 | 0.10 | 0.11 | 0.08 | 2.24 |
| Gd | 0.39 | 0.45 | 0.25 | 1.31 | 0.22 | 0.46 | 0.37 | 0.36 | 0.23 | 5.7 |
| Tb | 0.08 | 0.11 | 0.07 | 0.25 | 0.04 | 0.06 | 0.08 | 0.06 | 0.04 | 0.85 |
| Dy | 0.54 | 0.62 | 0.35 | 1.66 | 0.28 | 0.39 | 0.52 | 0.53 | 0.27 | 4.8 |
| Ho | 0.11 | 0.11 | | 0.31 | | 0.08 | | | | 0.87 |
| Er | 0.34 | 0.43 | 0.22 | 1.13 | 0.18 | 0.33 | 0.39 | 0.33 | 0.20 | 2.4 |
| Tm | 0.06 | 0.06 | 0.04 | 0.13 | 0.03 | 0.04 | 0.05 | 0.05 | 0.03 | 0.33 |
| Yb | 0.42 | 0.41 | 0.26 | 1.23 | 0.22 | 0.36 | 0.44 | 0.34 | 0.26 | 1.8 |
| Lu | 0.06 | 0.09 | 0.05 | 0.17 | 0.03 | 0.07 | 0.04 | 0.04 | 0.02 | 0.28 |
| Hf | 0.26 | 0.38 | 0.14 | 0.69 | 0.14 | 0.29 | 0.28 | 0.17 | 0.12 | 4.8 |
| Ta | 0.04 | 0.06 | 0.07 | 0.17 | 0.02 | 0.04 | −0.06 | 0.04 | 0.03 | 1.4 |
| W | 0.11 | 0.02 | | | | | | | | 0.15 |

Notes. AOA-22c and AOA-22p are separately digested samples of AOA-22 (see Samples and Methods for details). Element concentration data omitted are below the limit of detection, or, in the case of Mo and W affected by contamination from the drill bits. The USGS reference material BHVO-2 was analyzed alongside the AOA samples to monitor measurement accuracy; for most elements the data agree with the recommended values of Jochum et al. (2016) to within 15 %. Mineralogy: ol = olivine, aug/di = Al-Ti-rich augite/diopside, met = Fe-Ni-metal, sf = Fe-Ni-sulfide, an = anorthite, rt = rutile, spl = spinel, chr = chromite, px = Ca,Fe-rich pyroxene, nph = nepheline, sdl = sodalith.

iCAP TQ quadrupole inductively coupled plasma mass spectrometer (ICP-MS) at the Max Planck Institute for Solar System Research (MPS). Samples were introduced into the mass spectrometer in a 0.5 M $HNO_3$–0.01 M HF solution, and all elements were measured in SQ mode. Element concentrations of the samples were determined relative to an in-house multi-element standard solution. To monitor the accuracy of the measurements, the USGS reference sample BHVO-2 was analyzed alongside the AOA samples and for most elements the concentrations obtained match the recommended values of Jochum et al. (2016) to within better than ∼15 %, while the Mn and Cr concentrations agree to within 5 % (Table 1).

For high-precision Ti and Cr isotope measurements, Ti and Cr were separated from the remaining sample solutions following our established procedures (Gerber et al., 2017; Schneider et al., 2020). Titanium isotopic compositions were measured on a ThermoScientific Neptune *Plus* MC-ICP-MS at the Institut für Planetologie at medium mass resolution ($\Delta m/m > 4000$). Analyte solutions of ∼300 ng/g Ti were introduced using a Savillex C-Flow PFA concentric nebulizer connected to a Cetac Aridus II desolvating system, resulting in an ion beam intensity of ∼$3.1 \times 10^{-10}$ A on $^{48}$Ti. Measurements comprised two consecutive lines of data acquisition, where in the first line ion beams on all Ti isotope masses were measured in blocks of 40 cycles of 4.2 s integration time each. In this line, ion beams on masses 51 and 53 were also monitored, to correct for possible isobaric interferences of $^{50}$V and $^{50}$Cr on $^{50}$Ti. In the second line, all Ti isotopes were measured again in blocks of 20 cycles of 4.2 s integration time each, but this time together with the ion beam on mass 44, which was used to correct for isobaric interferences of $^{46}$Ca and $^{48}$Ca. The Ca interference correction was optimized by measuring

Ca-doped Ti solutions and manually adjusting the $^{46}$Ca/$^{44}$Ca and $^{48}$Ca/$^{44}$Ca ratios used for the correction in each session (Zhang et al., 2011). Polyatomic interferences (e.g., $^{22}$Ne$_2^+$ on mass 44, $^{36}$Ar$^{14}$N$^+$ on mass 50, and $^{40}$Ar$^{13}$C$^+$ on mass 53) were resolved by measuring on resolved flat top peak sections. Instrumental mass bias was corrected relative to $^{49}$Ti/$^{47}$Ti = 0.749766 (Niederer et al., 1981) using the exponential law. The Ti isotope measurements of the samples were bracketed by measurements of the OL-Ti standard (Millet and Dauphas, 2014), and the Ti isotopic compositions are reported in the ε-notation relative to this standard [$\varepsilon^i$Ti = ($^i$Ti/$^{47}$Ti)$_{sample}$/($^i$Ti/$^{47}$Ti)$_{OL-Ti}$− 1) × 10$^4$ where $i$ = 46, 48, 50]. The accuracy and reproducibility of the Ti isotope measurements were assessed by repeated analyses of BHVO-2 (Table 2), and results are in good agreement with previously published results for this sample (e.g., Gerber et al., 2017). Analytical uncertainties are reported as 95% confidence intervals.

Chromium isotopic compositions were measured on the Thermo-Scientific Triton *Plus* TIMS at the Institut für Planetologie. Rhenium filaments were loaded with 1 μg Cr together with an emitter solution consisting of silica gel mixed with H$_3$BO$_3$ and Al. Chromium isotopes were measured in multi-static mode using four lines where the ion beams on masses 50, 51, 52, and 53 were collected in the axial cup in lines 1 to 4, respectively. A single Cr isotopic measurement consisted of 10 blocks of static four-line measurement, where each block consisted of 20 cycles with 8.4 s integration time each. Baselines were measured every second block. Instrumental mass fractionation was corrected relative to $^{50}$Cr/$^{52}$Cr = 0.051859 (Shields et al., 1966) using the exponential law. The final $^{53}$Cr/$^{52}$Cr and $^{54}$Cr/$^{52}$Cr ratios were calculated as the mean fractionation-corrected $^{53}$Cr/$^{52}$Cr ratio from lines 1–4, and the mean fractionation-corrected $^{54}$Cr/$^{52}$Cr from lines 1–3 (due to the setup of the Faraday cups, $^{54}$Cr could not be measured in line 4). Possible isobaric interferences from $^{54}$Fe on $^{54}$Cr were corrected by also measuring $^{56}$Fe in the first line. This multi-static procedure led to improvement of the external reproducibility of our measurements to ±0.11 $\varepsilon^{53}$Cr and ±0.22 $\varepsilon^{54}$Cr (2 s.d.). The Cr isotopic compositions are reported in the ε-notation relative to the mean composition of the NIST SRM3112a Cr standard measured in each session [$\varepsilon^i$Cr = ($^i$Cr/$^{52}$Cr)$_{sample}$/($^i$Cr/$^{52}$Cr)$_{SRM3112a}$− 1) × 10$^4$ where $i$ = 53, 54]. The accuracy and reproducibility of the Cr isotope measurements were further assessed by repeated analyses of the USGS reference sample DTS-2b, and results for this sample are in good agreement with those obtained in a prior study (Schneider et al., 2020). Finally, analytical uncertainties are reported as 95 % confidence intervals (for $n \geq 4$) or 2 s. d. of the NIST SRM 3112a standard for the respective session (for $n < 4$).

## 3. Results

### 3.1. Chemical composition

Bulk concentration data for the AOAs of this study are provided in Table 1 and plotted in Fig. 1. The individual AOAs exhibit homogeneous enrichments in refractory elements relative to CI chondrites (Lodders, 2003), ranging from ~1.4 × CI in AOA-10, −18, and −23 to ~7 × CI in AOA-17. More volatile elements are depleted in AOAs, although Na, Rb, and Zn display modest enrichments compared to less volatile elements like Mn. This enrichment most likely results from secondary alteration of Allende on the parent body (Ruzicka et al., 2012). Another notable feature of the AOA's chemical composition is the depletion of siderophile elements relative to lithophile elements of similar condensation temperature; this metal-silicate fractionation possibly results from the physical separation of metal during the assembly of AOAs (Ruzicka et al., 2012). Overall, the bulk chemical composition of the AOAs of this study are consistent with that reported for AOAs in previous studies (e. g., Ruzicka et al., 2012; Krot et al., 2004), indicating that the samples investigated here are typical AOAs.

For two samples of this study, AOA-10 and AOA-23, the optical inspection of the drill holes suggests that these samples may be

**Table 2**
O, Ti, and Cr isotopic data for AOAs and two USGS reference materials.

| Sample | n (O) | $\delta^{18}$O | 2σ | $\delta^{17}$O | 2σ | $\Delta^{17}$O | 2σ | n (Ti) | $\varepsilon^{46}$Ti | 2σ | $\varepsilon^{48}$Ti | 2σ | $\varepsilon^{50}$Ti | 2σ | n (Cr) | $\varepsilon^{53}$Cr | 2σ | $\varepsilon^{54}$Cr | 2σ |
|---|---|---|---|---|---|---|---|---|---|---|---|---|---|---|---|---|---|---|---|
| AOA-1 | 5 | −40.27 | 6.24 | −43.26 | 5.91 | −22.32 | 2.73 | 15 | 1.39 | 0.09 | 0.18 | 0.03 | 7.42 | 0.12 | 18 | −0.18 | 0.09 | 5.64 | 0.11 |
| AOA-5 | 4 | −35.10 | 4.78 | −37.54 | 5.39 | −19.29 | 2.93 | 15 | 1.73 | 0.07 | 0.33 | 0.03 | 10.12 | 0.09 | 9 | 0.43 | 0.05 | 5.79 | 0.19 |
| AOA-10 | 3 | −46.23 | 1.03 | −47.63 | 0.99 | −23.59 | 0.47 | 10 | 0.52 | 0.14 | 0.26 | 0.05 | 6.14 | 0.05 | 3 | 0.26 | 0.19 | 3.49 | 0.22 |
| AOA-17 | 4 | −44.70 | 1.66 | −46.18 | 2.06 | −22.93 | 1.21 | 15 | 1.46 | 0.08 | −0.15 | 0.07 | 7.07 | 0.05 | 6 | 0.23 | 0.06 | 5.26 | 0.18 |
| AOA-18 | 3 | −46.97 | 0.83 | −48.13 | 0.90 | −23.71 | 0.52 | 15 | 1.35 | 0.09 | 0.27 | 0.11 | 7.31 | 0.08 | 5 | −0.05 | 0.11 | 3.74 | 0.19 |
| AOA-20 | 4 | −41.48 | 1.72 | −43.13 | 1.92 | −21.56 | 1.12 | 12 | 1.62 | 0.07 | 0.43 | 0.03 | 8.84 | 0.09 | 6 | 0.08 | 0.03 | 4.72 | 0.13 |
| AOA-22c | 3 | −43.97 | 1.55 | −45.43 | 1.01 | −22.57 | 0.24 | 9 | 1.61 | 0.09 | 0.37 | 0.05 | 9.01 | 0.09 | | | | | |
| AOA-22p | | | | | | | | 10 | 1.58 | 0.08 | 0.50 | 0.08 | 8.58 | 0.08 | 2 | 0.31 | 0.12 | 4.37 | 0.25 |
| AOA-23 | 3 | −43.03 | 1.17 | −44.47 | 0.95 | −22.09 | 0.44 | 12 | 1.23 | 0.05 | 0.27 | 0.05 | 6.72 | 0.11 | 9 | 0.09 | 0.06 | 1.45 | 0.15 |
| BHVO-2 | | | | | | | | 11 | −0.06 | 0.24 | −0.18 | 0.45 | −0.10 | 0.27 | | | | | |
| DTS-2b | | | | | | | | | | | | | | | 8 | 0.05 | 0.06 | 0.08 | 0.2 |

Notes. $n$ = number of individual isotope measurements. $\delta^i$O = [($^i$O/$^{16}$O)$_{sample}$/($^i$O/$^{16}$O)$_{SMOW}$ −1] × 10$^3$; where $i$ = 17, 18; $\Delta^{17}$O = $\delta^{17}$O − 0.52 × $\delta^{18}$O. $\varepsilon^i$Ti = [($^i$Ti/$^{47}$Ti)$_{sample}$/($^i$Ti/$^{47}$Ti)$_{OL-Ti}$ −1] × 10$^4$, where $i$ = 46, 48, 50. $\varepsilon^i$Cr = [($^i$Cr/$^{52}$Cr)$_{sample}$/($^i$Cr/$^{52}$Cr)$_{NIST\ SRM\ 3112a}$ −1] × 10$^4$, where $i$ = 53, 54. Uncertainties on Ti and Cr isotope data are 95% confidence intervals (for $n \geq 4$) or 2 s.d. of the standard (for $n < 4$ and for BHVO-2 and DTS-2b).

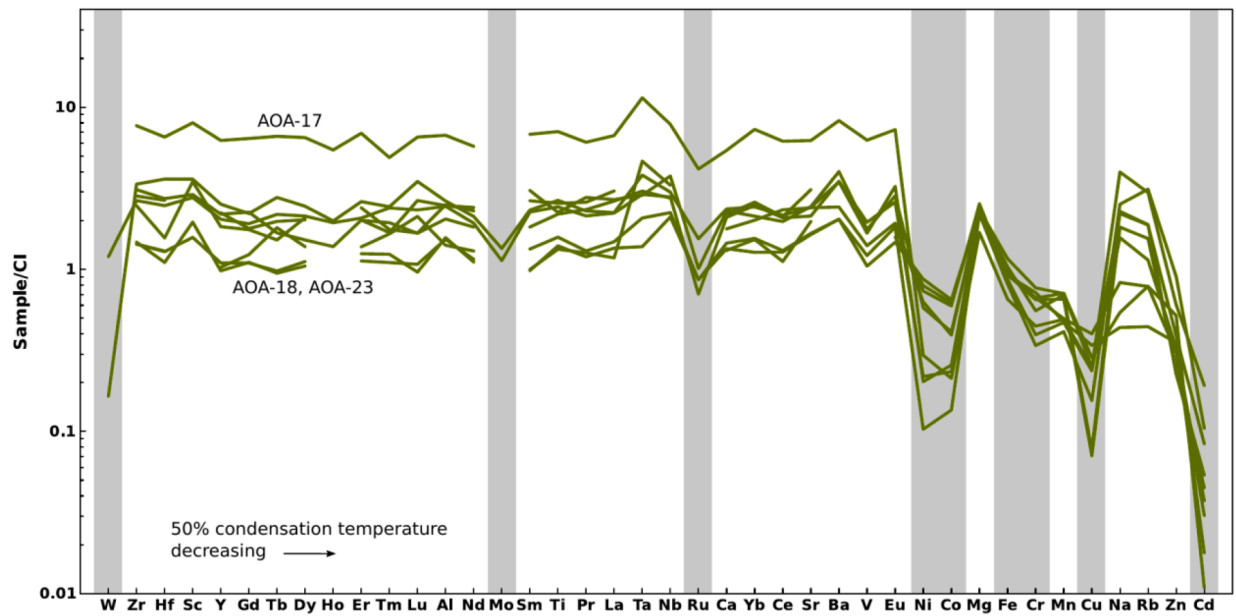

**Fig. 1.** CI chondrite-normalized elemental abundances for AOAs. Elements are plotted left to right in order of decreasing 50 % condensation temperature. The AOAs are enriched in refractory lithophile elements with overall flat abundance patterns, but are depleted in siderophile elements (marked with gray bars) and volatile elements. The depletion in siderophile elements reflects the paucity of metal in AOAs (Ruzicka et al., 2012). Barium in AOA-10 and AOA-22c is overprinted by secondary effects and not plotted.

contaminated with surrounding material (see above). Consistent with this, these two samples together with AOA-18 show lower enrichments (~1.3–1.6 × CI) in refractory lithophile elements than the other AOAs of this study (~2.2–7.1 × CI). Thus, some of the chemical variability among the AOAs of this study may be due to contamination with bulk Allende material during sample extraction. On the other hand, the strong enrichment of AOA-17 (~7.1 × CI) is attributable to the presence of relict CAI material, consistent with the optical identification of CAI material in the core of this sample during the initial SEM characterization (see above).

### 3.2. O, Ti, and Cr isotopes

The O isotope data for olivine in the AOAs of this study are reported in Table 2 and plotted in Fig. 2. The O isotopic compositions of all samples are relatively homogeneous and range from $\delta^{18}O = -37.5 \pm 5.4$ ‰ (2 s.d.) and $\delta^{17}O = -35.1 \pm 4.8$ ‰ (2 s.d.) for AOA-5 to $\delta^{18}O = -47.0 \pm 0.8$ ‰ (2 s.d.) and $\delta^{17}O = -48.1 \pm 1.9$ ‰ (2 s.d.) for AOA-18 (Table 2). The mean $\Delta^{17}O$ of the AOA olivines is $-22.3 \pm 2.8$ ‰ (2 s.d.), consistent with O isotope data for AOAs from previous studies (e.g., Krot et al., 2004). In O three-isotope space, the AOA olivines thus plot far off the terrestrial fractionation (TF) line and bulk Allende, along the primitive chondrite mineral (PCM) line in the same region as primitive, unaltered CAI minerals (see summary in Krot, 2019) (Fig. 2a).

All AOAs of this study reveal well-resolved and correlated anomalies in $\varepsilon^{46}Ti$ and $\varepsilon^{50}Ti$, ranging from ~0.52 to ~1.7 and ~6 to ~10, respectively (Table 2, Fig. 2b). The variations in $\varepsilon^{48}Ti$ are smaller and barely resolved, with values ranging from –0.15 to 0.50. As for O isotopes, the range in Ti isotope anomalies is indistinguishable from that observed for CAIs (e.g., Davis et al., 2018), which in turn is distinct from bulk Allende and all non-carbonaceous (NC) and carbonaceous (CC) meteorites (Fig. 2b). A similar observation is made for the $^{54}Cr$ anomalies, where $\varepsilon^{54}Cr$ values of the AOAs range from ~1.5 to ~6 (Table 2), which except for a single sample (AOA-23, $\varepsilon^{54}Cr = 1.45$) is again the same range of values as observed for CAIs (e.g., Torrano et al., 2023; Birck and Allegre, 1985; Fig. 2c).

### 3.3. $^{53}Mn$-$^{53}Cr$ systematics

In addition to the $^{54}Cr$ variations, the AOAs of this study also show variations in $^{53}Cr$, with $\varepsilon^{53}Cr$ values ranging from –0.18 to 0.43. Prior studies have shown that $\varepsilon^{53}Cr$ variations among meteorites and meteorite components predominately reflect decay of short-lived $^{53}Mn$ (e.g., Trinquier et al., 2008; Qin et al., 2010). This is consistent with the data of this study, given that there is no correlation between the $^{53}Cr$ and $^{54}Cr$ anomalies. Moreover, the $\varepsilon^{53}Cr$ values of the AOAs are broadly correlated with their $^{55}Mn/^{52}Cr$ ratios (Fig. 3); although the spread in the latter is limited and, thus, no isochron can be constructed, this broad trend is consistent with a predominantly radiogenic origin of the observed $^{53}Cr$ variations. Also shown in Fig. 3 are reference lines of the expected $\varepsilon^{53}Cr$ variations for Mn-Cr fractionation at $t = 0$ Ma and $t = 3$ Ma after CAI formation. Evidently, the AOAs of this study do not plot along a given reference line, indicating that the observed $\varepsilon^{53}Cr$ variations do not reflect a single event of Mn-Cr fractionation. Nevertheless, most of the $\varepsilon^{53}Cr$ variations are consistent with a relatively early Mn-Cr fractionation among the AOAs within the first Ma after CAI formation.

### 4. Discussion

#### 4.1. Common heritage of AOAs and CAIs

A key observation from the new data of this study is that AOAs and CAIs share the same nucleosynthetic isotope signatures for Ti and Cr, suggesting a close genetic link. While this has been shown for O isotopes previously (Hiyagon and Hashimoto, 1999; Krot et al., 2004), our study demonstrates that the link between AOAs and CAIs extends to nucleosynthetic isotope anomalies. However, before interpreting the isotopic data in terms of genetic relationships among samples, it is useful to assess how processes on the parent body and contamination during extraction of the AOAs may have modified the isotopic signatures observed. For instance, the primary mineralogy and O isotopic composition of AOAs from Allende may have been subject to change by thermal metamorphism and aqueous alteration on the parent body (e.g., Krot et al., 1995; Imai and Yurimoto 2003). This is evident from the occurrence of secondary minerals such as nepheline and sodalite in the AOAs of this study, suggesting that for instance the observed variable and high Na and Rb contents in these samples likely reflects transference of these elements into the AOAs during formation of these secondary minerals during parent body alteration. Nevertheless, in spite of these secondary modifications, the O isotope composition of the AOAs as obtained by analyzing forsteritic olivine almost certainly is a primary signature. This is because aqueous alteration on the parent body only produces Fe-rich olivine, and the diffusion rate of O in olivine is much lower than the diffusion rate of Fe (Chakraborty, 2010).

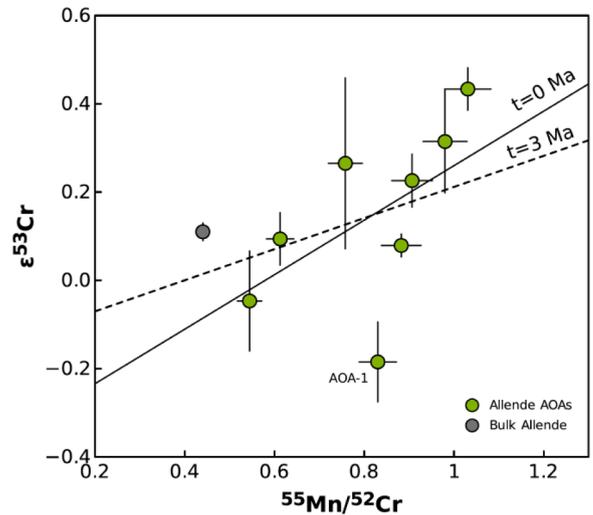

**Fig. 3.** $\varepsilon^{53}$Cr versus $^{55}$Mn/$^{52}$Cr for AOAs. Two reference isochrons corresponding to Mn-Cr fractionation at the time of CAI formation ($t = 0$ Ma) and 3 Ma later are shown. These reference lines were calculated using the mean composition of AOAs from this study ($^{55}$Mn/$^{52}$Cr = 0.815; $\varepsilon^{53}$Cr = 0.15) and a solar system initial $^{53}$Mn/$^{55}$Mn = $7 \times 10^{-6}$ (Tissot et al., 2017). Although the AOAs define no isochron, the $^{53}$Cr variations among most of the AOAs are consistent with an early Mn-Cr fractionation within the first few Ma, suggesting that most of the $\varepsilon^{53}$Cr variations among the AOAs are radiogenic in origin. The limited spread in Mn/Cr ratios and $^{53}$Cr isotopic compositions may in part reflect modifications by parent body processes and/or contamination with bulk Allende material during sample extraction (see text for details). Bulk Allende is shown for reference (Trinquier et al., 2008).

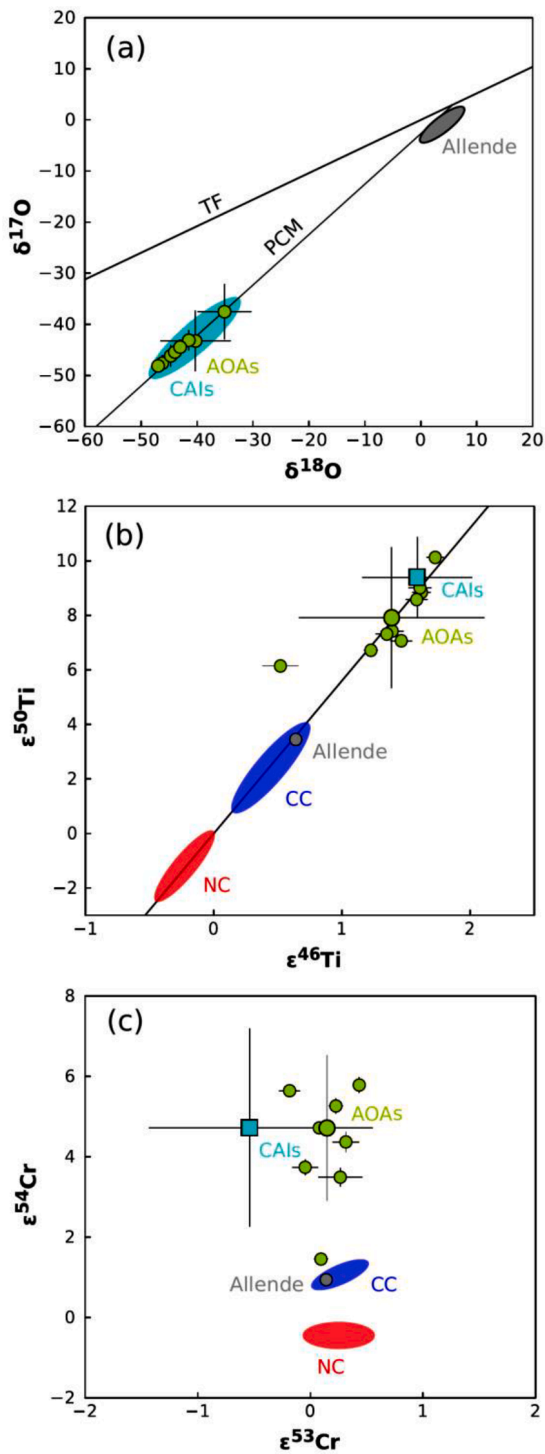

**Fig. 2.** O, Ti, and Cr isotopic composition for AOAs in comparison to CAIs, bulk Allende, and NC and CC meteorites. (a) The O isotope composition of AOA olivines overlaps with the composition of primitive CAIs and plot along the primitive chondrite mineral (PCM) line far from the bulk Allende host chondrite. (b-c) AOAs display strongly positive $\varepsilon^{46}$Ti, $\varepsilon^{50}$Ti, and $\varepsilon^{54}$Cr values, and are displaced from the NC and CC meteorite fields. The average Ti and Cr isotopic composition of AOAs (large green circles) overlap with the average compositions of CAIs. The only exception is AOA-23, which plots close to the composition of bulk Allende, most likely because this sample has been contaminated with bulk Allende material during sample extraction. NC and CC fields, and CAI and Allende isotopic compositions based on data compilations in Burkhardt et al. (2019), Spitzer et al. (2020), Zhu et al. (2021), Torrano et al. (2019; 2023), and Krot (2019).

Several observations indicate that the Ti isotopic compositions of the AOAs have not been modified significantly by parent body processes or through the contamination with surrounding bulk Allende material during sample extraction. First, Ti and other refractory lithophile elements occur in broadly chondritic relative proportions in the AOAs of this study. This observation holds regardless of whether an element is mobile during parent body processes (e.g., Sr) or immobile (e.g., Ti), indicating there has been limited exchange of refractory elements like Ti between AOAs and the surrounding sample matrix. Second, the Ti isotopic compositions of all AOAs is vastly different from that of bulk Allende, but similar to that of CAIs. Yet, on average CAIs contain ~5 times and AOAs ~2 times more Ti than bulk Allende, and so if there were significant mobilization and exchange of Ti during parent body processes, we would expect a gradual change of the Ti isotopic composition of AOAs away from CAIs and towards the composition of bulk Allende. In a plot of $\varepsilon^{50}$Ti versus 1/Ti, the AOAs indeed plot along a linear trend suggestive of mixing between $^{50}$Ti- and Ti-rich with $^{50}$Ti- and Ti-poorer materials, but this trend passes well above the composition of bulk Allende and Allende matrix (Fig. 4a). As such, the $\varepsilon^{50}$Ti variations among the AOAs of this study cannot result from simple addition of or exchange with bulk Allende material, but most likely reflect the heterogeneous distribution of CAIs within the AOAs. This is consistent with petrographic evidence for relict CAI material in AOAs, and with the observation that many CAIs plot on the extension of the $\varepsilon^{50}$Ti versus 1/Ti trend defined by the AOAs (Fig. 4a). Together, these observations indicate that the chemical and isotopic inventory of refractory elements like Ti in the AOAs of this study have not been significantly modified by processes on the parent body or by contamination during sample extraction.

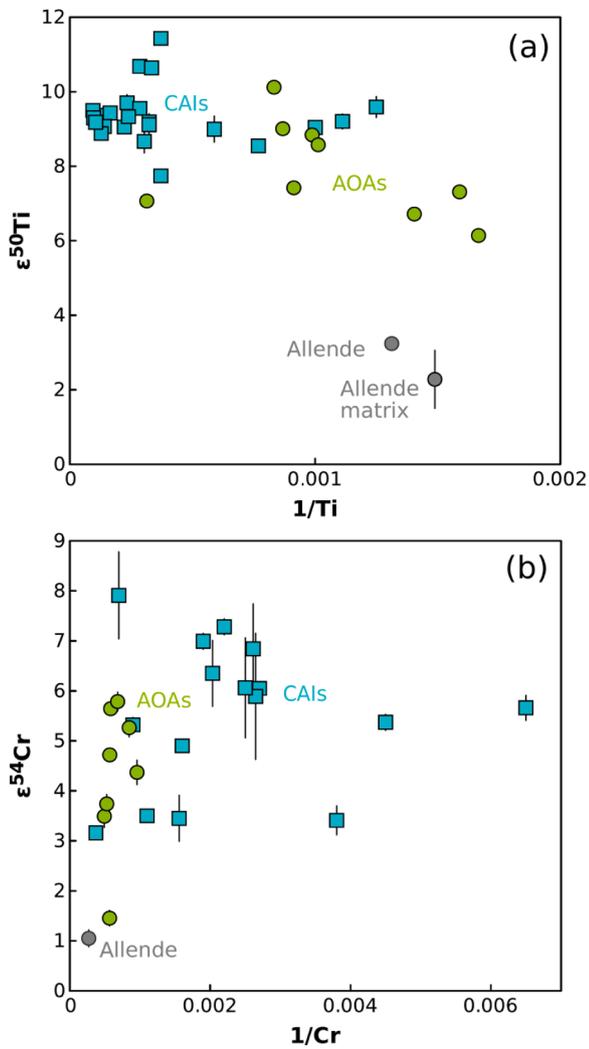

**Fig. 4.** Relation of Ti and Cr isotopic compositions and concentrations in AOAs and CAIs. (a) In an $\varepsilon^{50}$Ti versus 1/Ti diagram AOAs define a weak trend, suggestive of mixing between $^{50}$Ti- and Ti-rich and $^{50}$Ti- and Ti-poorer materials. This trend passes well above the composition of bulk Allende and Allende matrix, demonstrating that $\varepsilon^{50}$Ti variations among the AOAs cannot reflect simple addition of bulk Allende material. Instead, the $\varepsilon^{50}$Ti variations most likely reflect the heterogeneous incorporation of CAIs, which plot on the extension of the $\varepsilon^{50}$Ti versus 1/Ti trend. CAI data from Torrano et al. (2019); Allende data from Gerber et al. (2017). (b) The $\varepsilon^{54}$Cr variations among the AOAs do not define a single two-component mixing line, but the lower $\varepsilon^{54}$Cr of some AOAs trend towards the composition of bulk Allende, suggesting some modifications occurred during parent body processes and/or by contamination with bulk Allende material during sample extraction. CAIs have even lower Cr concentrations (i.e., higher 1/Cr) than AOAs and some of the $\varepsilon^{54}$Cr variability among the CAIs may also be due to secondary modifications. CAI data from Birck and Allegre (1985) and Torrano et al. (2023), data for bulk Allende from Schneider et al. (2020).

For Cr the situation is more complex, mostly because AOAs (and CAIs) are depleted in Cr relative to bulk Allende, making the Cr isotopic composition of AOAs (and CAIs) more susceptible to modifications. This is evident from plots of $\varepsilon^{53}$Cr versus $\varepsilon^{54}$Cr (Fig. 2c) and $\varepsilon^{54}$Cr versus 1/Cr (Fig. 4b), where the Cr isotope variations among the AOAs extend towards the composition of bulk Allende. For instance, AOA-23, for which we estimate that up to ~50 % contamination during sample extraction may have occurred (see above), plots close to the composition of bulk Allende, suggesting this sample is strongly contaminated. Sample AOA-10, which may have been subject to a similarly high contamination during extraction as AOA-23, also has a somewhat lower $\varepsilon^{54}$Cr of ~3.5 than most other AOAs from this study (Fig. 2c). However, AOA-18 displays a similar low $\varepsilon^{54}$Cr of ~3.8, but for this sample we have no clear indication for contamination during sample extraction. One possibility is that the Cr isotopic composition of this sample has been modified during parent body alteration, but this is difficult to assess with certainty. Taken together, in plots of $\varepsilon^{54}$Cr versus $\varepsilon^{53}$Cr (Fig. 2c) and 1/Cr (Fig. 4b) the three AOAs with the lowest $\varepsilon^{54}$Cr trend towards the composition of bulk Allende, and so for these samples the original Cr isotopic signatures may well have been modified by parent body processes and/or contaminated during sample extraction. For most samples, inspection of the drill holes reveals that contamination is less than 10 %, which given the approximately twofold higher Cr concentration of bulk Allende compared to the AOAs may lower an AOA's $\varepsilon^{54}$Cr by ~1, which is similar to the range of $\varepsilon^{54}$Cr values among the AOAs of this study (excluding AOA-10 and −23). Addition of bulk Allende material to some of the AOAs may also account for the limited spread in Mn/Cr ratios of these samples and the observation that the samples with the most extreme radiogenic $\varepsilon^{53}$Cr compositions are also those with the largest $\varepsilon^{54}$Cr anomalies (Fig. 2c). This does not mean that there are no indigenous $\varepsilon^{54}$Cr variations among AOAs, but due to the possibility of contamination during sample extraction and modification during parent body alteration, this cannot be said with confidence. As such, the original $\varepsilon^{54}$Cr of AOAs is probably best approximated by the samples with the largest anomalies.

In summary, the O and Ti isotopic compositions of the AOAs of this study seem to predominantly reflect their original signatures. As such, we find identical $^{16}$O-rich composition of primary unaltered minerals in AOAs and CAIs, and indistinguishable mean $\varepsilon^{50}$Ti values of 8.1 ± 2.6 and 9.4 ± 1.5 for AOAs and CAIs, respectively (all uncertainties 2 s.d., CAI data from Torrano et al., 2019; 2023). For Cr, the isotopic compositions of some of the AOAs may have been modified by parent body processes or during extraction of the samples. Nevertheless, the average $\varepsilon^{54}$Cr of AOAs of 4.7 ± 1.8 (2 s.d., excluding AOA-23) is indistinguishable from the average $\varepsilon^{54}$Cr = 4.7 ± 2.5 (2 s.d.) of CAIs (Torrano et al., 2019, 2023). To this end, it is important to recognize that owing to their low Cr contents the Cr isotope signatures of CAIs may have been similarly modified by parent body processes as those of AOAs. Further, both AOAs and CAIs exhibit similar maximum $\varepsilon^{54}$Cr values of ~6–7 (Torrano et al., 2023; this study), and so even if all of the $\varepsilon^{54}$Cr variability of AOAs and CAIs were due to secondary modification, both would still have indistinguishable indigenous $^{54}$Cr signatures.

The combined O, Ti, and Cr isotopic data of this study, therefore, indicate formation of AOAs and CAIs from a common isotopic reservoir. As AOAs formed at lower temperatures (~1200–1350 K) than CAIs (>1300 K), this reservoir must have maintained a constant isotopic composition over this temperature range. Moreover, given the vastly different isotopic composition of bulk meteorites, chondrules, and matrix compared to CAIs and AOAs (Figs. 5, 6), the isotopic composition of solid matter in the disk must have changed after the formation of CAIs and AOAs.

*4.2. Origin of isotope anomalies in CAIs and AOAs*

The finding of a common isotopic composition for CAIs and AOAs has important implications for understanding the origin of the isotope anomalies in these materials. There are two competing models which have been proposed to account for the isotope anomalies in CAIs. In one model, the anomalies reflect thermal processing of the precursor material of the CAIs (Paton et al., 2013; Trinquier et al., 2009). This model is in part motivated by chemical and mineralogical evidence for formation of CAIs from a gas of solar composition, which itself formed by evaporation of material close to the Sun. Thus, the formation of CAIs themselves is evidence for extreme thermal processing and it thus is tempting to associate the vastly different isotopic composition of CAIs compared to later formed solids with this thermal processing. In the other, competing model the isotopic anomalies in CAIs are unrelated to thermal processing, but reflect the isotopic composition of the early disk, which was later modified by late infall of material with distinct isotopic composition (Nanne et al., 2019; Burkhardt et al., 2019; Jacquet et al., 2019). The new isotopic data for AOAs from this study help to distinguish between these two disparate models.

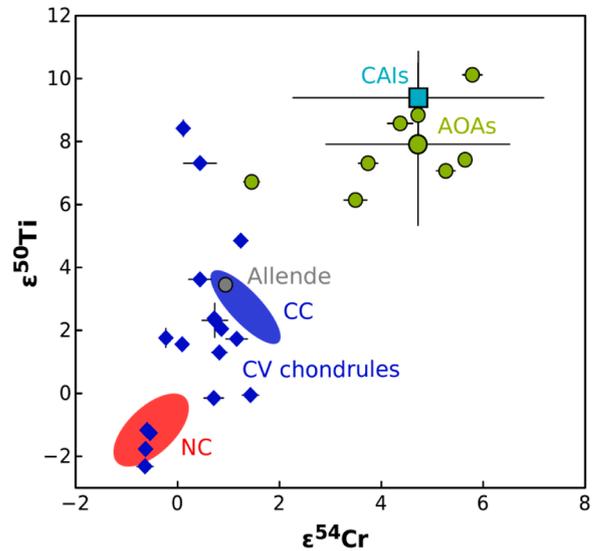

**Fig. 6.** Same as Fig. 5c, but now also including data for individual CV chondrules (Schneider et al., 2020; Williams et al., 2020). The heterogeneous distribution of AOAs and CAIs among chondrule precursors can account for the observed variations in $\varepsilon^{50}$Ti and $\varepsilon^{54}$Cr among the CV chondrules. However, on average the chondrules have much lower $\varepsilon^{50}$Ti and $\varepsilon^{54}$Cr values than AOAs, indicating that AOAs cannot be the main precursor to chondrules. Instead, AOAs and chondrules formed from different populations of dust having distinct Cr and Ti isotopic compositions.

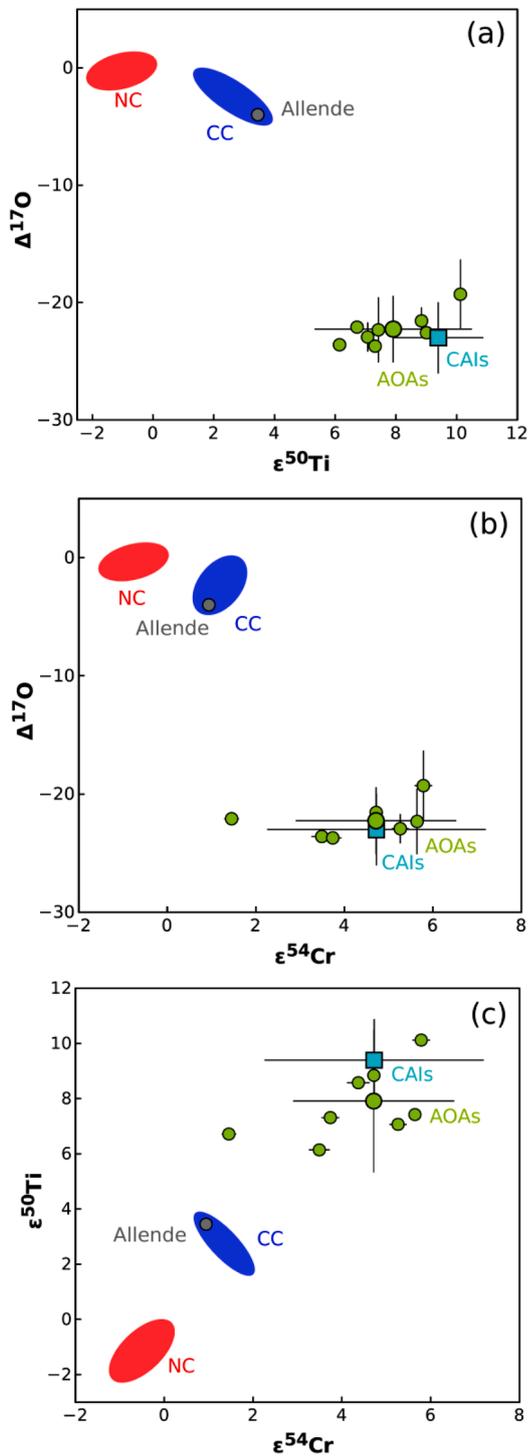

**Fig. 5.** Isotope anomalies in O, Ti, and Cr in multi-element space. In all three plots, AOAs and average CV3 CAIs have the same isotopic compositions, which is far offset from the compositions of bulk NC and CC meteorites. CC meteorites consistently plot between the NC field and the composition of CAIs/AOAs, suggesting the CC composition results from mixing between materials with NC- and CAI/AOA-like isotopic compositions. NC and CC fields, and average CAI and Allende isotopic compositions based on compilations in Burkhardt et al. (2019), Spitzer et al. (2020), Zhu et al. (2021), Torrano et al. (2019, 2023), and Krot (2019).

Thermal processing may produce isotopic heterogeneity in two different ways: either isotopically anomalous grains are preferentially evaporated to produce an isotopically anomalous gas from which CAIs and AOAs later condense; or thermally more resistant, isotopically anomalous grains became enriched by removal of thermally more labile materials, before they are themselves evaporated to produce the CAI- and AOA-forming gas. For instance, Trinquier et al. (2009) argued that thermally labile presolar silicates enriched in $^{50}$Ti and $^{46}$Ti were preferentially evaporated into the CAI-forming gas; however, such carriers have not been identified in meteorites. Instead, primitive chondrites contain $^{54}$Cr- and $^{50}$Ti-rich oxide grains, which are suggested carriers of $^{54}$Cr and $^{50}$Ti anomalies in meteorites and their components (Dauphas et al., 2010; Qin et al., 2011; Nittler et al., 2018). These oxide grains are refractory and, as such, expected to be thermally stable during evaporation in the solar nebula. Consequently, if these grains are the carriers of Cr and Ti isotope anomalies in CAIs and AOAs, generating these anomalies by thermal processing would require the prior removal of thermally less stable materials to produce a dust reservoir enriched in $^{54}$Cr and $^{50}$Ti, which is subsequently evaporated to produce the gas from which CAIs and AOAs later condensed.

Regardless of the presolar carriers involved, our finding of a common isotopic reservoir for CAIs and AOAs is not easily reconciled with a thermal processing origin of the isotope anomalies. The formation of CAIs is thought to involve vaporization of dust that is injected into the solar system close to the Sun (i.e., inside of the CAI condensation line), followed by outward spreading and cooling of the gas, which allows condensation of the gas to form CAIs. As noted above, AOAs formed at lower temperature than CAIs, meaning that they formed further away from the Sun. Consequently, the common isotopic composition of CAIs and AOAs requires that the isotopic composition of the gas did not change during this outward spreading, i.e., its composition was not modified by interaction with the ambient gas and dust of the disk. However, as the AOA condensation line marks a lower temperature in the disk than the CAI condensation line, any thermal processing at the location of the AOA condensation line had been less severe. Thus, in this scenario, the isotopic composition of the outward flowing gas would likely be modified by interaction with the ambient disk, and so AOAs should have a less anomalous isotopic composition than CAIs, in particular for non-refractory elements like Cr. This can only be circumvented if the gas quickly spread beyond the centrifugal radius (i.e., the radius at which material from the molecular cloud is falling onto the disk), such that there would be no pre-existing disk with which the precursor material of the AOAs could mix and exchange. But in this case, there is no reason why the isotope anomalies of CAIs should reflect thermal processing, simply because at this early stage in disk evolution, before formation of a disk, all material was falling close to the Sun where it likely was completely vaporized, leaving little room for producing isotope anomalies by selective thermal processing.

Another, more straightforward way to avoid dilution of the AOA's isotopic signature by interaction with the ambient disk is that the disk simply had the same isotopic composition as CAIs and AOAs. In an attempt to account for the isotopic difference between NC and CC meteorites, Nanne et al. (2019) and Burkhardt et al. (2019) proposed that CAIs record the isotopic composition of early infalling material, which also is the material that largely built the initial disk surrounding the forming Sun. In this model, the distinct isotopic composition of later formed solids, as recorded in the isotopic composition of bulk meteorites, reflects a change in the isotopic composition of the infalling matter over time. Models of the infall process show that any isotopic heterogeneity within the protosolar cloud is likely imparted onto the disk, because material from different areas of the cloud fall onto different areas of the disk (Dauphas et al., 2002; Visser et al., 2009; Jacquet et al., 2019). This infall model naturally accounts for the common isotopic composition of CAIs and AOAs, because this composition simply reflects the composition of the initial disk. Thus, unlike in the thermal processing case, there is no isotopic difference between more and less strongly processed areas of the disk, and so the interaction of AOAs with the ambient disk does not alter their isotopic composition.

In summary, although thermal processing seems an attractive option to generate isotope anomalies in CAIs and AOAs, in detail this model has difficulty in accounting for the common isotopic composition of these refractory inclusions. By contrast, this common composition is a natural outcome of the heterogeneous infall model, in which the isotopic composition of CAIs/AOAs reflects the composition of the early disk. As such, the new data of this study lend strong support to the idea that the nucleosynthetic isotope heterogeneity among and between CAIs/AOAs and bulk meteorites is inherited from the Sun's parental molecular cloud, and was imparted onto the disk during infall. One important implication of this model is that the isotopic composition of CAIs/AOAs likely closely resembles that of the Sun, while CI chondrites, which provide the closest chemical match to the Sun (Lodders, 2003), are isotopically distinct. This is because the Sun accreted most of its mass during the early and main infall phase, and so the late-infalling material had a profound impact only on the isotopic composition of the disk, but not on the isotopic composition of the Sun itself. This is consistent with the indistinguishable O isotope compositions of CAIs/AOAs and the Sun (McKeegan et al., 2011).

In addition to CAIs and AOAs, there are several other refractory inclusions, including platy hibonite crystals (PLACs), spinel-hibonite inclusions (SHIBs), and FUN (fractionated and unidentified nuclear effects) CAIs (Dauphas and Schauble, 2016). Together, these refractory inclusions are characterized by a much larger range of isotope anomalies and lower initial $^{26}$Al/$^{27}$Al ratios than the 'normal' CAIs we have discussed so far. These characteristics have been interpreted to indicate an earlier formation of these more anomalous inclusions, prior to the injection of $^{26}$Al into the solar system (Liu et al., 2009; Köop et al., 2016). The much wider range of isotope anomalies in the earlier formed refractory inclusions may then reflect the progressive homogenization of the isotope heterogeneities, from the most anomalous PLACs over the less anomalous SHIBs to the normal CAIs, which show less variable and smaller isotope anomalies than the PLACs and SHIBs (e.g., Köop et al., 2016; Render et al., 2019; Torrano et al., 2023). Within the framework of the model presented here, the range of isotope anomalies among earlier-formed refractory inclusions suggests that the early infalling material was heterogeneous itself, and that normal CAIs (and AOAs) represent the average isotopic composition of this material.

*4.3. Origin of isotopic difference between refractory inclusions and chondrule precursors*

While the overall chemical composition of AOAs is more refractory than those of chondrules, the dominant mineral in AOAs is olivine, raising the question of whether AOA olivine is the precursor to olivine in Type I chondrules. On the basis of trace element data, Ruzicka et al. (2012) showed that chondrule olivine could have formed by melting of AOA olivine and subsequent olivine-melt fractionation during chondrule crystallization. These authors also noted that the distinct O isotope compositions of AOAs and chondrules requires open-system gas-melt interaction to produce the more $^{16}$O-rich compositions of the chondrules. Along similar lines, the generally Si-poor composition of AOAs compared to chondrules requires the addition of Si to the chondrule melts (Krot et al., 2004; Marrocchi et al., 2019). As such, from a chemical and O isotope point of view, chondrules could have been derived by melting of AOA precursors, followed by gas-melt chemical and isotopic exchange (e.g., Ruzicka et al., 2012).

The Cr and Ti isotopic data of this study can be used to test this idea. These isotope anomalies represent the composition of the precursor dust of AOAs and chondrules and unlike for O isotopes, are not expected to be modified significantly by gas-melt interaction [although the temperatures during chondrule formation can exceed the condensation temperatures of most elements (Hewins and Radomsky, 1990), any Ti and Cr in the gas phase will stem from the solid precursors]. Prior studies have shown that the heterogeneous distribution of AOAs and CAIs among chondrule precursors can readily account for the observed variations in $\varepsilon^{50}$Ti and $\varepsilon^{54}$Cr values among carbonaceous chondrite chondrules (Gerber et al., 2017; Schneider et al., 2020; Williams et al., 2020). However, as shown in Fig. 6, on average these chondrules have much lower $\varepsilon^{50}$Ti and $\varepsilon^{54}$Cr values than AOAs. As such, AOAs cannot represent the main precursor material of the chondrules, and instead AOAs and chondrules on average must have formed from different populations of nebular dust having distinct Cr and Ti isotopic compositions.

A corollary of this observation is that despite the petrologic and chemical evidence for a continuous evolution of nebular condensates from CAIs to AOAs to chondrule precursors (e.g., Ruzicka et al., 2012), there is a disconnect between the precursor material of CAIs and AOAs on the one side, and of chondrules on the other. This disconnect appears to be related to a change in the isotopic composition of solid material in the disk from relatively $^{50}$Ti- and $^{54}$Cr-rich to $^{50}$Ti- and $^{54}$Cr-poorer compositions. As argued above, this change is unlikely to reflect thermal processing of CAI/AOA precursors, because in this case, AOAs would be expected to be isotopically intermediate between CAIs and chondrules. By contrast, this change can readily be understood within the framework of the heterogeneous infall model, where the isotopic composition of infalling matter during collapse of the protosolar cloud changes over time from the $^{50}$Ti- and $^{54}$Cr-rich composition measured in CAIs/AOAs to the $^{50}$Ti- and $^{54}$Cr-poor composition of non-carbonaceous (NC) meteorites. It is noteworthy that individual chondrules from the CV chondrite Allende cover almost this entire range of compositions, including the characteristic $^{50}$Ti- and $^{54}$Cr-poor composition of NC meteorites as well as compositions more enriched in $^{50}$Ti and $^{54}$Cr than even the carbonaceous chondrites (Fig. 6). These variations most likely reflect isotopic heterogeneity among the chondrule precursors, which itself reflects mixing of NC material with variable amounts of CAIs/AOA-like materials. The isotopic heterogeneity among chondrites can, therefore, be understood as providing a record of the temporal evolution of condensate isotope compositions from CAI/AOA-like to NC-like compositions, followed by transport and mixing of these materials in the disk.

Fig. 7 is a cartoon illustrating our preferred scenario for the isotopic evolution of the solar nebula during infall, and the subsequent mixing of nebular condensates having different isotopic compositions. This scenario is based on the model of Nanne et al. (2019) for the origin of the NC—CC dichotomy. In this scenario, CAIs form close to the Sun by evaporation of infalling matter from the protosolar cloud, followed by viscous outflow and cooling of the gas, upon which first CAIs and then AOAs condense at their respective condensation lines. Both objects are transported outwards through the rapid radial expansion of the infalling material, which is the reason why they are predominantly found in carbonaceous chondrites (e.g., Ciesla, 2007). In the inner disk, infall continues but now its composition changes to NC; high-temperature condensates still form close to the Sun at this stage, but now have an NC-like isotopic composition. Evidence for this is found, for instance, in Na-Al-rich chondrules from ordinary chondrites, which contain a refractory, CAI-like material with an NC-like $^{50}$Ti isotope signature (Ebert et al., 2018). At this stage, material is also infalling onto the disk further away from the Sun and as a result is not completely vaporized. This material is the dominant precursor of carbonaceous chondrite chondrules, which, since it formed in the inner disk, has an NC-like isotopic composition. Mixing of these different components with their distinct isotopic compositions then gave rise to the isotopic heterogeneity observed among chondrule precursors. Later melting of these materials during chondrule formation did not modify their nucleosynthetic isotope signatures, and so the isotopic diversity among solar nebula condensates is preserved in the composition of the chondrules.

Finally, we note that the same mixing processes likely also produced the distinct isotopic compositions of NC and CC meteorites. Evidence for this comes from the observation that CC meteorites generally have isotope compositions intermediate between those of NC meteorites and CAIs/AOAs, implying that the CC reservoir contains a greater proportion of isotopically CAI/AOA-like material than the NC reservoir (Nanne et al., 2019; Burkhardt et al., 2019; Render et al., 2022). We, therefore, propose that mixing between solar nebula condensates, which acquired their distinct isotopic compositions as a result of time-varied infall from a heterogeneous protosolar cloud, is the dominant and overarching process producing the nucleosynthetic isotope heterogeneity of the solar nebula.

**CRediT authorship contribution statement**

**Christian A. Jansen:** Writing – original draft, Methodology, Investigation, Formal analysis. **Christoph Burkhardt:** Writing – original draft, Visualization, Validation, Supervision, Project administration,

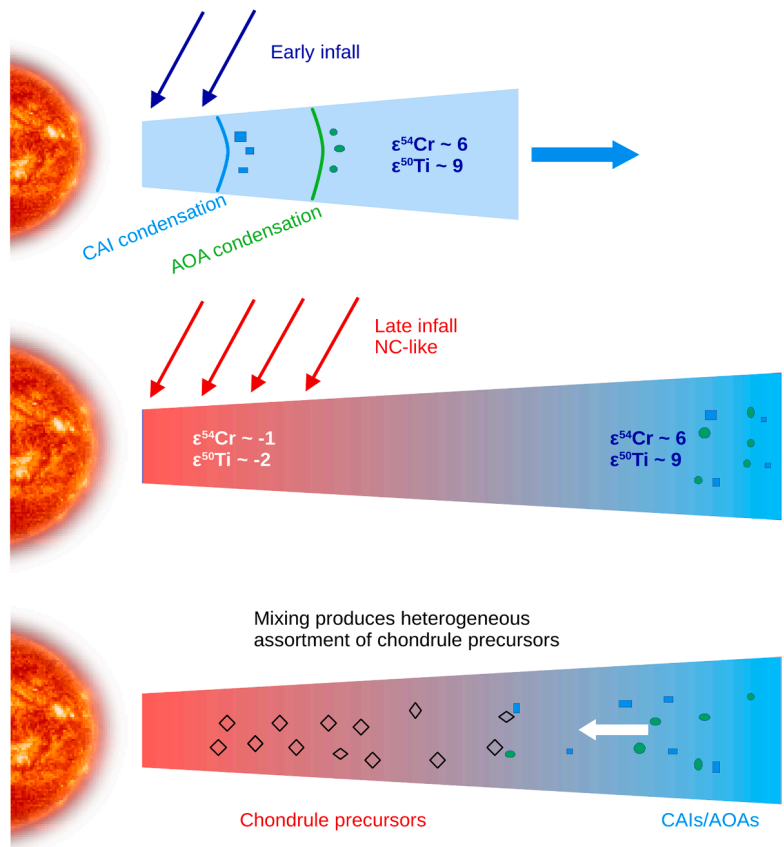

**Fig. 7.** Cartoon illustrating our preferred scenario for the isotopic evolution of the solar nebula during infall, and the subsequent mixing of nebular condensates having different isotopic compositions. Scenario is based on the models of Nanne et al. (2019); Burkhardt et al. (2019); and Jacquet et al. (2019) for the origin of the NC—CC dichotomy. CAIs form close to the Sun by evaporation of infalling matter from the protosolar cloud, followed by viscous outflow and cooling of the gas, upon which first CAIs and then AOAs condense. Both objects are transported outwards through the rapid radial expansion of the disk. In the inner disk, infall continues but now its composition changes to NC; high-temperature condensates still form close to the Sun at this stage, but now have an NC-like isotopic composition. At this stage, material is also infalling onto the disk further away from the Sun and as a result is not completely vaporized. This material is the dominant precursor of carbonaceous chondrite chondrules, which, since it formed in the inner disk, has an NC-like isotopic composition. Mixing of these different components with their distinct isotopic compositions then gave rise to the isotopic heterogeneity observed among chondrule precursors.

Investigation, Formal analysis, Conceptualization. **Yves Marrocchi:** Writing – review & editing, Validation, Resources, Methodology, Investigation, Formal analysis. **Jonas M. Schneider:** Writing – review & editing, Methodology, Formal analysis. **Elias Wölfer:** Writing – review & editing, Methodology, Formal analysis. **Thorsten Kleine:** Writing – review & editing, Validation, Supervision, Resources, Investigation, Funding acquisition, Conceptualization.

**Declaration of competing interest**

The authors declare that they have no known competing financial interests or personal relationships that could have appeared to influence the work reported in this paper.

**Data availability**

Data will be made available on request.

**Acknowledgments**

We thank K. Metzler for assistance with the SEM analyses, two anonymous reviewers for helpful comments, and F. Moynier for efficient editorial handling. This project has received funding from the European Union's Horizon 2020 research and innovation program under grant agreement No. 871149 (Europlanet).

**Supplementary materials**

Supplementary material associated with this article can be found, in the online version, at doi:10.1016/j.epsl.2024.118567.